\documentclass[nofootinbib,preprint,superscriptaddress,showpacs,preprintnumbers,amsmath,amssymb,aps,pre]{revtex4-2}

\usepackage[utf8]{inputenc} 
\usepackage{footnote}
\usepackage{epsfig}
\usepackage{natbib}
\usepackage{amsmath}
\usepackage{times}
\usepackage{epstopdf} 
\usepackage{graphicx}
\usepackage{color}
\usepackage{placeins}
\usepackage[ruled,vlined]{algorithm2e}

\usepackage{afterpage}

\begin{document}

\title{Chaotic resetting: A resetting strategy for deterministic chaotic systems}

 \author{Julia Cantisán}
  \email[]{julia.cantisan@dci.uhu.es }
 \affiliation{Grupo de Física No Lineal (FQM-280), Departamento de Ciencias Integradas y Centro de Estudios Avanzados en Física, Matemáticas y Computación, Universidad de Huelva, 21071 Huelva, Spain}

\author{Alexandre R. Nieto}
 \affiliation{Nonlinear Dynamics, Chaos and Complex Systems Group, Departamento de
Biología y Geología, Física y Química Inorgánica, Universidad Rey Juan Carlos, Tulip\'{a}n s/n, 28933 M\'{o}stoles, Madrid, Spain}

\author{Jes\'{u}s M. Seoane}
 \affiliation{Nonlinear Dynamics, Chaos and Complex Systems Group, Departamento de
Biología y Geología, Física y Química Inorgánica, Universidad Rey Juan Carlos, Tulip\'{a}n s/n, 28933 M\'{o}stoles, Madrid, Spain}

\begin{abstract}

Restarting a stochastic search process can accelerate its completion by providing an opportunity to take a more favorable path with each reset. This strategy, known as stochastic resetting, is well studied in random processes. Here, we introduce {\it chaotic resetting}, a fundamentally different resetting strategy designed for deterministic chaotic systems. Unlike stochastic resetting, where randomness is intrinsic to the dynamics, chaotic resetting exploits the extreme sensitivity to initial conditions inherent to chaotic motion: unavoidable uncertainties in the reset conditions effectively generate new realizations of the deterministic process. This extension is nontrivial because some realizations may significantly speed up the search, while others may significantly slow it down.  We study the conditions required for chaotic resetting to be consistently advantageous, concluding that it requires the presence of a mixed phase space in which fractal and smooth regions coexist. We quantify its effectiveness by demonstrating substantial reductions —ranging from $40\%$ to $90\%$— in average search times when an optimal resetting interval is used. These results establish a clear conceptual bridge between deterministic chaos and search optimization, opening new avenues for accelerating processes in real-world chaotic systems where perfect control or knowledge of initial conditions is unattainable.

\end{abstract}

\maketitle
\newpage

\section{Introduction} \label{sec:Introduction}

Stochastic resetting (SR) is an effective strategy to speed up a stochastic search process, based on the idea that restarting can accelerate completion by repeatedly offering the system new opportunities to follow more favorable paths. The conceptual origins of SR can be traced back to mathematical models developed in the 1970s, particularly in the context of population dynamics, where population resets were considered for optimal pest control \cite{Kyriakidis1989}. Later, resetting protocols were used to improve the performance of randomized algorithms \cite{Luby1993}. A comprehensive historical overview of these early developments is provided in \cite{Maso2023}. However, it was not until 2011 with the influential work of Evans and Majumdar \cite{Evans2011} that stochastic resetting was formally introduced and the term was coined. One of the most striking results of their study is that, when stochastic resetting is applied, the time required for a diffusing particle to locate a target in an infinite domain becomes finite. Furthermore, this acceleration is optimized at a specific resetting rate, for which the mean search time is minimized.

Since then, SR has been extensively extended from different perspectives: exploring multiple targets \cite{Bressloff2020}, multiple searchers \cite{Bhat2016}, active searchers \cite{Kumar2020}, nondiffusive processes \cite{Kusmierz2014, Masoliver2019}, or searchers confined to a potential \cite{Singh2020, Cantisan2021, Cantisan2024}, to name a few. Recently, interesting applications of SR have been developed, such as the enhanced sampling of molecular dynamics simulations \cite{Blumer2022} or applying stochastic resetting to physical problems, as the Ising model \cite{Magoni2020}, or quantum mechanics \cite{Yin2023, Yin2024}. See \cite{Evans2020} for a detailed review. 

Previous studies on resetting have focused exclusively on stochastic systems, as resetting to the same (or nearly the same)  initial state in a purely deterministic system would provide no advantage: the same (or nearly the same) trajectory and search time would simply be reproduced. However, this limitation does not apply to chaotic systems. Chaotic systems are deterministic dynamical systems characterized by extreme sensitivity to initial conditions: an infinitesimally small difference in initial conditions can lead to completely different system evolutions. In real-world systems, exact initial conditions cannot be perfectly reproduced or even measured. Furthermore, this is also true for numerical computations, where round-off errors are inevitable. In chaotic systems, these small differences grow exponentially over time, causing trajectories that start arbitrarily close to each other to eventually diverge dramatically. The smaller the initial difference, the longer it takes for the divergence to become noticeable. However, given enough time, even the slightest change in initial conditions leads to significant deviations. With this in mind, resetting could serve as an effective strategy to accelerate a chaotic search process.

Here, we develop a strategy for resetting in chaotic systems: chaotic resetting (CR). In this approach, each reset initiates a new trajectory with an initial condition slightly perturbed from the original by a small quantity, referred to as uncertainty. We investigate the conditions under which chaotic resetting is beneficial and analyze how search time scales with uncertainty. To demonstrate the effectiveness of this method, we apply it to a search process—specifically an escape from a potential—and compare escape times with and without resetting. For this purpose, we consider a chaotic scattering problem in the Hénon--Heiles potential, a system that provides an ideal setting for our method: chaos and a mixed phase space.

The problem of controlling the duration of chaotic processes has traditionally been addressed through active control strategies based on localized perturbations. For example, in Ref.~\cite{Lilienkamp2020}, tailored perturbations are applied to a spatially extended system, and those trajectories that evolve closest to a desired target state are selected. Related approaches have been used to regulate the collective dynamics of self-propelled particles by applying local accelerations \cite{Medeiros2024}, to reduce long-lasting transients  by intermittently perturbing to a calculated spatial control line \cite{Ray2021}, as well as to confine chaotic trajectories within prescribed regions of phase space (so-called safe sets) using the method of partial control \cite{Capeans2017}. In contrast to these control-based methodologies, the approach proposed here is significantly less invasive and requires substantially less detailed knowledge of the underlying dynamics, as it relies solely on restarting the process rather than on the continuous application of finely tuned perturbations.

This paper is organized as follows. In Sec.~II, we describe the system we employ to present the method, that is, the Hénon--Heiles system. In Sec.~III, we explore the conditions under which chaotic resetting is an effective strategy. In Sec.~IV, we implement the method and present the results. Later, in Sec.~V, we present some metrics to measure the efficacy of chaotic resetting. Finally, in Sec.~VI, we present our concluding remarks and discuss broader implications.

\section{Toy model: the Hénon--Heiles system}

To test the effectiveness of chaotic resetting, we have chosen the Hénon--Heiles Hamiltonian. Although this system was originally introduced to study stellar motion in galactic potentials, it has become a paradigmatic example in nonlinear dynamics due to the rich variety of dynamical behaviors that the system exhibits for different energy values \cite{Aguirre2001, Barrio2008}. Over time, it has found applications in quantum chaos \cite{Efthymiopoulos2006, Brack1993}, general relativity \cite{Vesely2000,Zotos2020}, and chemical physics \cite{Feit1984,Waite1981}, among others. Furthermore, the system has frequently been used as a testbed for numerical techniques \cite{Barrio2009}, chaos detection algorithms \cite{Hillebrand2022,Zotos2015}, control of chaos \cite{Coccolo2013,Nieto2022}, and machine learning algorithms \cite{EscobarRuiz2024}.

The Hénon--Heiles Hamiltonian is given by

\begin{equation}
	{\cal{H}}(x,y,\dot{x},\dot{y})=\frac{1}{2}(\dot{x}^2+{\dot{y}}^2)+\frac{1}{2}(x^2+y^2)+x^2y-\frac{1}{3}y^3,
\end{equation}
where $x$ and $y$ are the spatial coordinates, and $\dot{x}$ and $\dot{y}$ are the velocity components. The last three terms correspond to the Hénon--Heiles potential.
\begin{figure}[ht]
	\centering
   \includegraphics[clip,width=0.55\textwidth,trim=0cm 0cm 0cm 0cm]{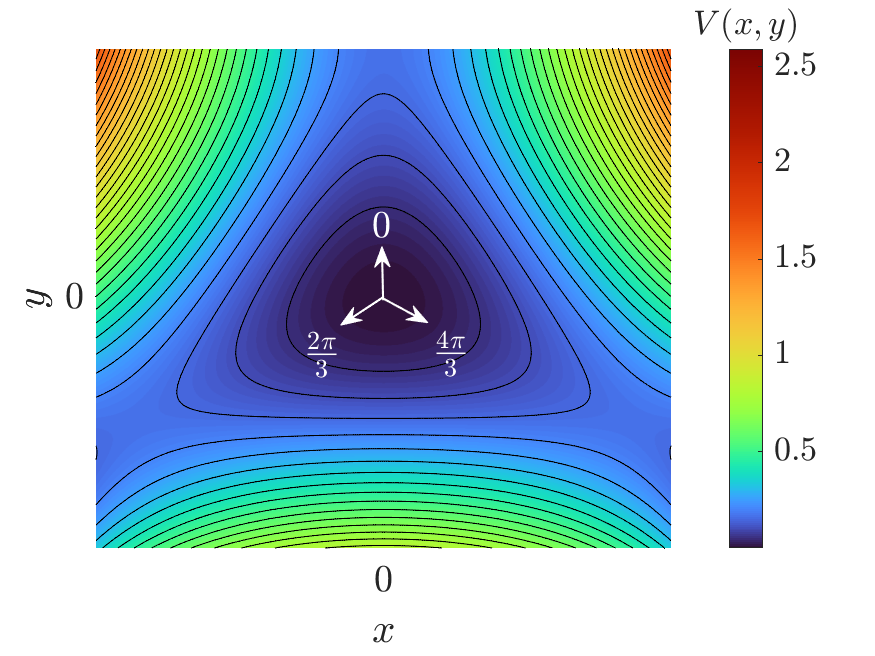}	
	\caption{Potential and isopotential curves for our toy model: the Hénon--Heiles system. For total energy values below the escape threshold $E_e = 1/6$, the isopotential curves are closed and the motion is confined. At energies $E > E_e$, the isopotential curves open, giving rise to three symmetric escape channels located at angles $\theta = 0,\, 2\pi/3,$ and $4\pi/3$. In this regime, trajectories can escape to infinity, and the system becomes a paradigmatic example of chaotic scattering.}
	\label{fig:1}
\end{figure}
Since the Hamiltonian has no explicit time dependence, the energy is a conserved quantity (i.e., ${\cal{H}}(x,y,\dot{x},\dot{y})=E$) and the equations of motion can be expressed as 
\begin{equation} \label{eq_motion}
	\begin{aligned}
		\ddot{x} & = -x - 2xy, \\
		\ddot{y} &= -y - x^2 + y^2.
	\end{aligned}
\end{equation}
For all numerical simulations in this work, these equations were integrated using a fourth-order Runge-Kutta method with a fixed time step $h=0.001$.

We represent the Hénon--Heiles potential in Fig.~\ref{fig:1} together with several isopotential curves. These curves indicate which regions of configuration space are accessible for a given value of $E$, since motion is restricted to the domain where the potential energy does not exceed $E$.
For energy values below the threshold $E_e=1/6$ trajectories are bounded, while for $E>E_e$ the potential features three symmetric exits, allowing particles to escape towards $\pm \infty$. In this regime, both regular (non-chaotic) and transient chaotic motion coexist. In this work, we focus on energies above $E_e$, where escapes are possible. This makes it possible to assess whether resetting can effectively reduce the escape time.

\begin{figure}[ht]
	\centering
    \includegraphics[clip,width=0.55\textwidth,trim=0cm 0cm 0cm 0cm]{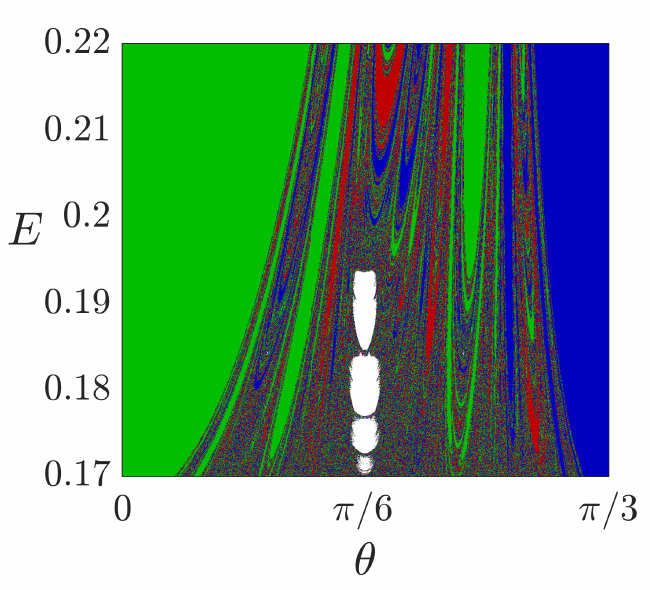}	
	\caption{Exit basins diagram of the Hénon--Heiles system in the $(\theta,E)$ plane. For a given energy, greater than $E_{e}$, particles are launched from the origin $(x,y)=(0,0)$ with an initial angle $\theta$.  The colors indicate the asymptotic behavior of the particle depending on the values of $E$ and $\theta$. The colors green, blue, and red denote the exit through which the particle escapes, corresponding to the upper, right, and left exits, respectively (see Fig.~\ref{fig:1}). White represents initial conditions that remain dynamically trapped within KAM tori and never escape.}
	\label{fig:2}
\end{figure}

In Fig.~\ref{fig:2}, we show the exit basins diagram for energies above $E_e$. The color of each region (red, green or blue) indicates the exit through which a given trajectory escapes. The white color indicates that the particle does not escape. All trajectories are launched from the origin $(x_0,y_0)=(0,0)$ with an initial angle $\theta$. Since the energy is conserved, the velocity is fixed as $(\dot{x}_0,\dot{y}_0)=(\sqrt{2E} \sin{\theta}, \sqrt{2E} \cos{\theta})$. As seen in Fig.~\ref{fig:2}, for all energy levels $E>E_e$, the system exhibits a mixed phase space where fractal regions associated with chaotic orbits coexist with smooth regions corresponding to regular orbits. The latter corresponds to initial conditions that lead to a fast escape. For example, a smooth region that spans a wide range of angles near $\theta=0$ indicates regular trajectories that escape quickly through the upper exit. Conversely, trajectories whose initial condition fall within the fractal regions experience transient chaos, resulting in long escape times. This occurs because these trajectories start near the fractal basin boundaries (i.e., the stable manifold of the chaotic saddle \cite{Aguirre2001}). In addition to the large smooth regions, other regions of the same nature can be found embedded in the fractal sea, but they are not noticeable due to the resolution of the figure. As we shall show later, the fact that the basin boundary does not occupy the entire phase space plays a crucial role in the effectiveness of chaotic resetting.

Additionally, within the considered energy range, the dynamics is more complex due to the existence of Kolmogorov-Arnold-Moser (KAM) islands~\cite{Ott}, which are depicted in white in Fig.~\ref{fig:2}. These islands correspond to non-chaotic trajectories that remain dynamically trapped within invariant tori and thus do not escape. These islands disappear at $E \approx 0.195$.

\section{Method viability} \label{sec:method_viability}

In this section, we investigate the conditions under which resetting may be a good strategy to reduce the search (escape) time in a chaotic system. We point to properties that are common across a wide range of systems, not just the one under consideration.

In stochastic search processes, the time $T$ to find the target is a random variable, often characterized by a heavy-tailed distribution. A useful metric to assess its variability is the coefficient of variation ($CV$), defined as the ratio of the standard deviation to the mean: $CV=\sigma(T)/\bar{T}$. A smaller $CV$ indicates lower variability relative to the mean, while a larger $CV$ suggests greater variability. It has been established that resetting can expedite a search process only if $CV>1$ \cite{Pal2017}. The reasoning behind this is that resetting interrupts long and inefficient search trajectories, replacing them with new attempts from the reset point. This truncation reduces the heavy tails of the first-passage time distribution, thereby lowering the variability of the search time. With this in mind, we now examine this metric in our problem. 

We introduce an uncertainty $\delta$ in the shooting angle $\theta$. Thus, we calculate the escape time for a trajectory that is launched from the origin with $\theta=\theta_0+\delta U$, where $U$ is a random variable that follows a uniform distribution ${\cal{U}}(-1,1)$. Thus, the distribution of escape times depends on both the uncertainty and the shooting angle. Without loss of generality, for the remainder of this work we fix the energy to $E=0.17$, which is slightly above $E_{e}$.

In the previous section, we analyzed the phase space structure and identified a rich dynamical landscape featuring fractal regions that exhibit transient chaos, smooth regions with fast escapes, and KAM islands. Intuitively, if a trajectory with a given $\theta_0$ exhibits transient chaos, but the $\delta$-neighborhood around $\theta_0$ includes a smooth region, the resulting escape time distribution will contain both long-lived and short-lived trajectories.  
 This variability can lead to $CV>1$, making resetting a promising strategy. Resetting gives the system the opportunity to access one of the trajectories that belong to the smooth region and escape fast. This is just one of several scenarios where chaotic resetting is beneficial.

\begin{figure}[h]
	\centering
	\includegraphics[clip,width=0.46\textwidth,trim=0cm 0cm 0cm 0cm]{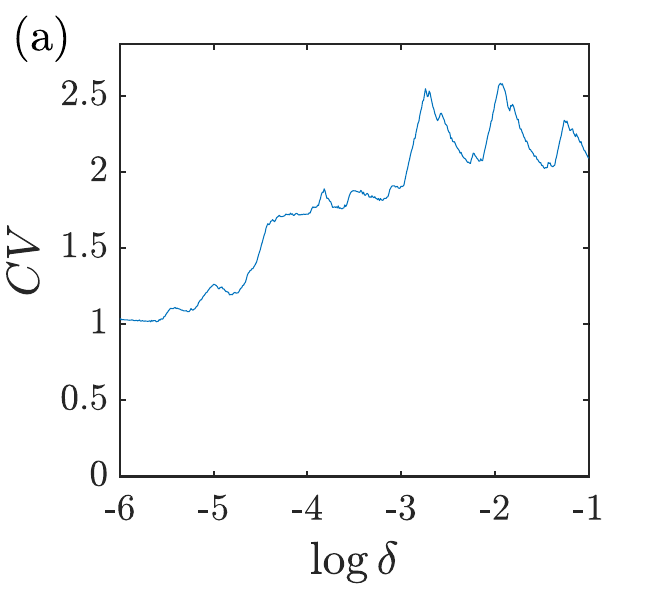}	
    \includegraphics[clip,width=0.46\textwidth,trim=0cm 0cm 0cm 0cm]{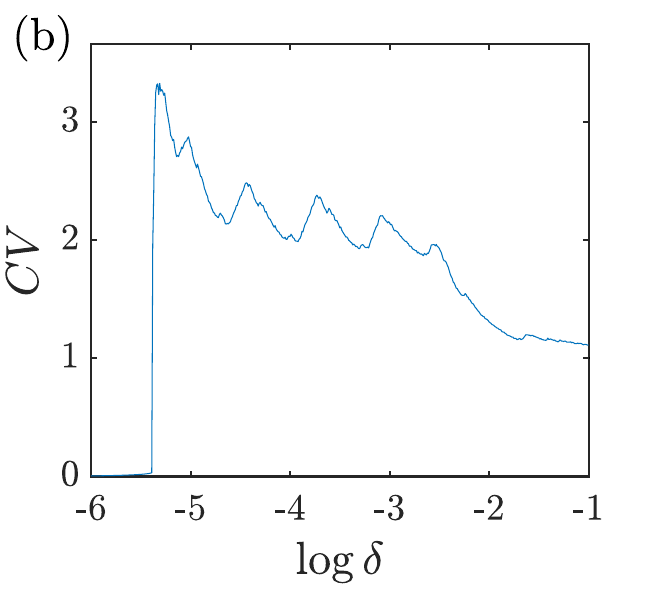}	
	\caption{Coefficient of variation (CV) as a function of the base-10 logarithm of the uncertainty \(\delta\) for two representative scenarios. In both cases, particles are launched with initial angles of the form \(\theta = \theta_0 + \delta U\), where \(U\) is a uniformly distributed random variable. For each value of \(\delta\), CV is computed from \(5 \times 10^5\) trajectories. Panel (a) shows the typical behavior for a particle launched with an angle \(\theta_0\) corresponding to a chaotic trajectory, i.e., an angle belonging to a fractal region of phase space (\(\theta_0=0.112\)). Panel (b) displays the typical behavior for a particle launched with an angle \(\theta_0\) from a smooth region of phase space (\(\theta_0=0.294\)). In both cases, CV exceeds unity when the \(\delta\)-neighborhood overlaps mixed regions of phase space containing both fractal and smooth structures. This occurs for \(\delta < 10^{-5}\), a range consistent with inherent uncertainties in physical systems, indicating that CR constitutes an effective strategy for reducing escape times.}  
	\label{fig:3}
\end{figure}

In Fig.~\ref{fig:3}(a), we represent $CV$ for $\theta_0=0.112$ as a function of the uncertainty $\delta$. This angle corresponds to a trajectory with transient chaotic motion. For small enough uncertainty $\delta$, we observe that $CV$ is only slightly larger than $1$. This occurs because the uncertainty is too small to reach any large smooth region, and the initial condition remains in a fractal region where two initially close trajectories present radically different escape times. Consequently, if a trajectory takes a long time to escape, resetting to a nearby initial condition may be advantageous. However, to determine whether resetting is consistently beneficial, we must consider the average behavior over all initial conditions within a given $\delta$-neighborhood. When performing this average, we find that the escape time distribution is roughly exponential, indicating that the system behaves locally as a hyperbolic chaotic system \cite{Burton2023}. In such systems, the standard deviation equals the mean, leading to $CV=1$. Thus, on average, resetting within the fractal region does not provide a systematic advantage. Therefore, if the system is fully chaotic (i.e., the basin boundary covers the entire phase space), chaotic resetting is ineffective. However, in most systems smooth and fractal regions coexist, introducing deviations from strict hyperbolicity. As a result, in most cases $CV$ is slightly greater than $1$.  This is precisely the present case, where narrow smooth regions are embedded in the chaotic sea. 

For larger $\delta$, the average escape time decreases while $CV$ increases, indicating that a smooth region has been reached. The successive peaks in $CV$ correspond to specific regions of phase space becoming accessible within the $\delta$-neighborhood of the chosen $\theta_0$. This behavior is characteristic of a trajectory that starts relatively close to the boundary of a smooth region. 

Another relevant scenario occurs when a trajectory starts in a smooth region. For values of $\delta$ smaller than the width of the region, all trajectories have the same escape time, leading to $\sigma(T)=0$. However, as $\delta$ increases, the fractal regions are reached and $CV$ increases. This behavior is captured in Fig.~\ref{fig:3}(b) for $\theta_0=0.294$. Notably, this transition occurs for $\delta<10^{-5}$, which is remarkably small relative to the characteristic scale of our system. This suggests that such an effect is not an artificially induced phenomenon, but rather an inherent and unavoidable consequence of natural uncertainties. For larger values of $\delta$, the fractal region occupies most of the $\delta$-neighborhood, leading to a perturbed hyperbolic regime in which $CV$ remains slightly above $1$.

To summarize, when uncertainty extends over a region of the phase space that includes both smooth and fractal regions, resetting can effectively prevent trajectories from remaining in fractal regions for long periods of time. This generally holds for systems with a mixed phase space. In Sec.~\ref{sec:efficacy}, we further investigate how commonly this scenario arises in our system and, more broadly, in chaotic systems.

Finally, the presence of KAM islands introduces a third possible scenario. If a KAM island falls within the $\delta$-neighborhood, resetting is counterproductive because these trajectories remain trapped. In this case, $CV$ becomes a meaningless measure as the average escape time tends to infinity. This, together with the fact that the presence of KAM islands is a property of the specific model we use, is the reason why we do not explore this scenario further in this paper. Notably, in the particular case in which the initial condition is in the KAM island, resetting is beneficial if the $\delta$-neighborhood falls out of the KAM island. 

\FloatBarrier
\section{Chaotic resetting}

In this section, we present the results of applying the chaotic resetting technique. Each time we reset the system, the initial conditions are given by $(x_0,y_0, \dot{x_0},\dot{y_0})=(0,0,\sqrt{2E} \sin{\theta}, \sqrt{2E} \cos{\theta})$, with the angle perturbed by $\theta=\theta_0 + \delta U$. Depending on the context in which this technique is applied, resetting can be understood either as repositioning the same particle to a new initial condition or as launching a new particle while discarding the previous one. Each approach has its own cost: the first involves a refractory period \cite{Evans2019, Sunil2023}, while the second implies the cost of launching a new trajectory.

\begin{figure}[h]
	\centering
	\includegraphics[clip,width=0.55\textwidth,trim=0cm 0cm 0cm 0cm]{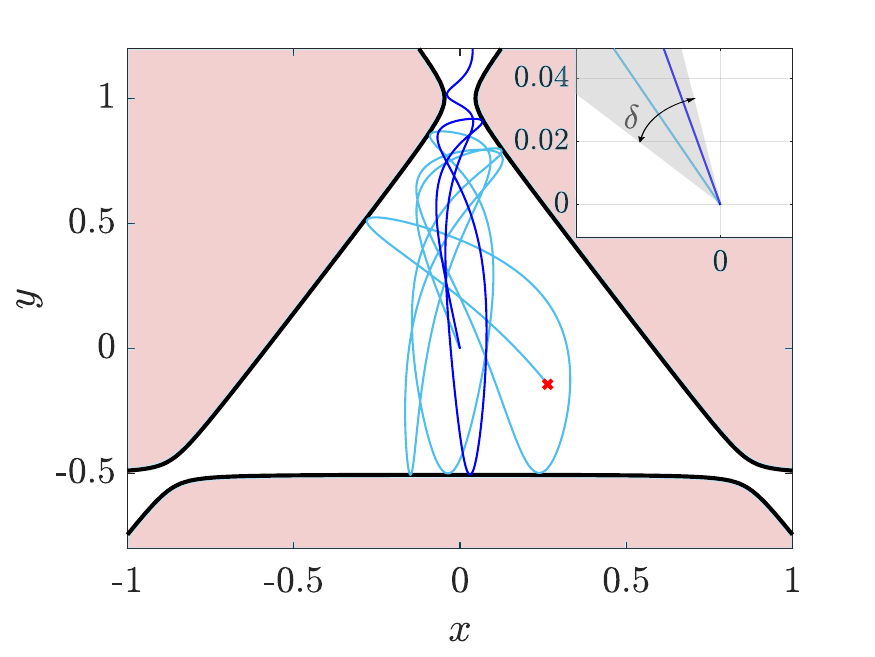}	
	\caption{Chaotic resetting in the Hénon--Heiles system. The light blue trajectory is launched from the origin with angle $\theta_0$ and evolves until it is stopped at a certain time $t_r$, marked by a red cross. The experiment is then repeated and a new (dark blue) trajectory is launched from the same initial position. Due to the physical impossibility of reproducing exactly the same launch angle, the new trajectory is launched with a slightly different angle $\theta=\theta_0+\delta U$ (as shown in the inset zoom). This new trajectory escapes quickly through the upper exit.}
	\label{fig:4}
\end{figure}

In Fig.~\ref{fig:4}, we illustrate a trajectory (light blue) that has not escaped by a given time $t_r$. Consequently, it is discarded at the position $(x(t_r),y(t_r))$, marked by the red cross in the figure. Subsequently, another trajectory (dark blue) starts with an angle perturbed by the uncertainty $\delta$ (depicted as the gray area in the inset). This second trajectory escapes before the resetting time $t_r$ is reached. Regarding resetting times, many time distributions have been explored in the literature \cite{Nagar2016, Bhat2016, Eliazar2020, Tal2025}. For simplicity, we focus on applying sharp restart, meaning that we reset at fixed time intervals.

\begin{figure}[ht]
	\centering
    \includegraphics[clip,width=0.5\textwidth,trim=0cm 0cm 0cm 0cm]{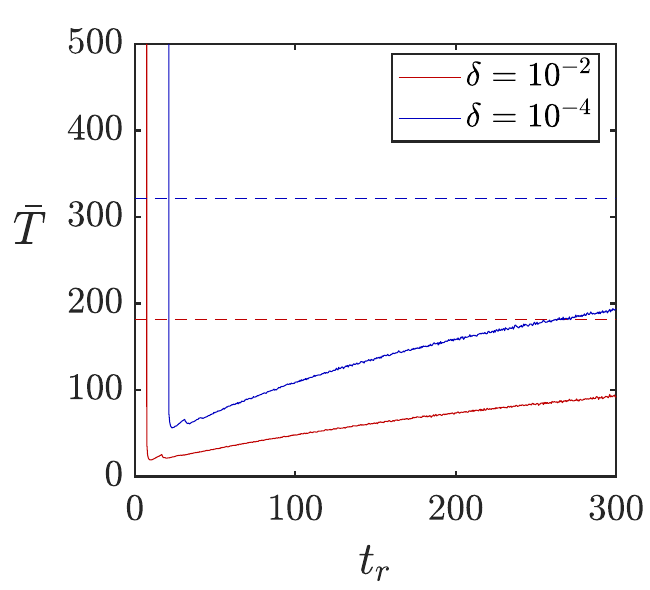}	
	\caption{Average escape time as a function of the resetting interval $t_r$ for two  uncertainty levels, $\delta=10^{-2}$ (red) and $10^{-4}$  (blue). The horizontal dashed lines represent the average escape time without resetting. The resetting strategy is advantageous whenever the average escape time falls below the dashed line of the same color, a condition that is satisfied over a broad range $t_r$ values. At the optimal resetting interval, the reduction in the mean escape time exceeds $80\%$ relative to the no-resetting case. For this simulation, a shooting angle corresponding to a chaotic trajectory  was used ($\theta_0=0.112$, same value as in Fig.~\ref{fig:3}), $600$ values of $t_r$ were considered, and $5\times10^4$ realizations were computed for each value of $t_r$. }
	\label{fig:5}
\end{figure}

We apply chaotic resetting and measure the average escape time, $\bar{T}$, for different values of the resetting time interval, $t_r$. The results are shown in Fig.~\ref{fig:5}. We present the dynamics for two different uncertainty values and compare the results with the average escape time without resetting, which in each case is indicated by a horizontal dashed line. We observe the typical behavior of stochastic resetting. For small values of $t_r$, the average escape time diverges because the system is reset before the fastest trajectory has had a chance to escape. For very large values of $t_r$, $\bar{T}$ approaches the average escape time without resetting. Between these extreme situations, there is a minimum that corresponds to the optimal resetting time interval $t_r^*$. For $\delta=10^{-2}$, $\bar{T}(t_r^*)$ shows an $89\%$ reduction, while for $\delta=10^{-4}$, $\bar{T}(t_r^*)$ shows an $83\%$ reduction. This reduction is calculated relative to the average escape time without resetting. To calculate this average, we consider initial conditions within a $\delta$-neighborhood, so the escape time depends on the value of $\delta$ ($\bar{T}=181.4$ for $\delta=10^{-2}$ and $\bar{T}=321.2$ for $\delta=10^{-4}$). In both cases, the reduction in escape time is impressive, and the minima occur very close to the undesired scenario where $\bar{T} \rightarrow \infty$. Given this behavior, in practical terms, it would be more reliable to choose a resetting time slightly larger than $t_r^*$. 

To aid reproducibility, in Appendix A we provide a pseudocode summarizing the chaotic resetting strategy used throughout this work. In particular, the results shown in Fig.~\ref{fig:5} can be reproduced by running the algorithm $5\times10^4$ times for $600$ values of $t_r$ from $0.5$ to $300$.

We have illustrated the results for a particular angle; however, during our research, we found similar results for many different angles with $CV>1$.

\FloatBarrier
\section{Efficacy of the resetting} \label{sec:efficacy}

In this section, we delve deeper into the efficacy of chaotic resetting and examine how it scales with uncertainty.

\subsection{Under what conditions is chaotic resetting beneficial?}

\begin{figure}[ht]
	\centering
    \includegraphics[clip,width=0.45\textwidth,trim=0cm 0cm 0cm 0cm]{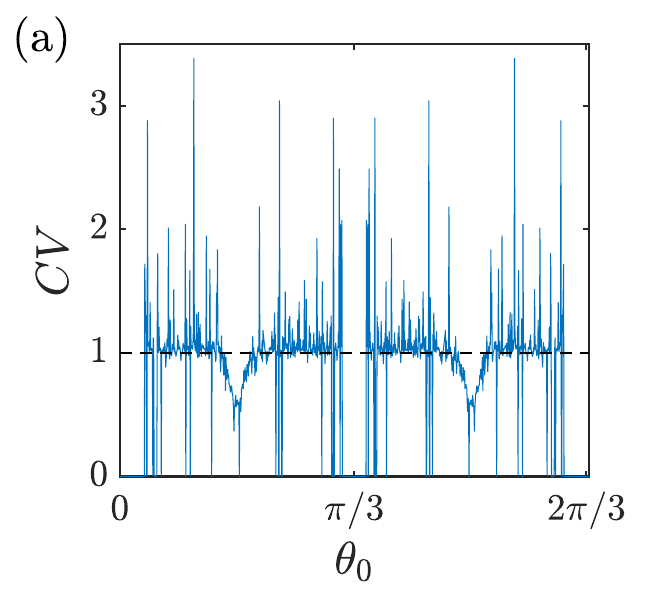}	
    \includegraphics[clip,width=0.45\textwidth,trim=0cm 0cm 0cm 0cm]{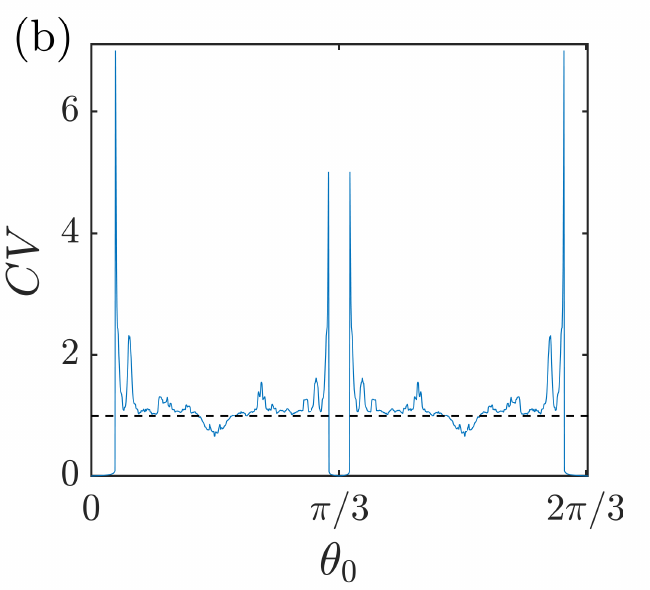}
	\caption{CV dependence on the shooting angle $\theta_0$ for uncertainty values (a)  $\delta=10^{-4}$ and (b) $\delta=10^{-2}$. For each angle $\theta_0$, the experiment is repeated $5\times10^5$ times, each time with a slightly different angle $\theta=\theta_0+\delta U$. Thus, for each angle $CV$ has been calculated using the escape time of these $5\times10^5$ shootings. The percentage of angles with $CV>1$, i.e., angles for which resetting would be an advantageous strategy, is $52\%$ for (a) and $72\%$ for (b). The curve in panel (b) is smoother as the $\delta$-neighborhood for each angle is larger and covers regions that are more similar, with less variability in escape times. Meanwhile, in panel (a), the $\delta$-neighborhood for a certain angle is narrower, causing that a $\delta$-neighborhood might include a smooth region, with trajectories that escape fast, and the adjacent $\delta$-neighborhood might overlap a region with long-lasting transients. This variability leads to the fractal-like structure observed in the $CV$ curve. In both panels, $1000$ different shooting angles $\theta_0$ were computed.}
	\label{fig:6}
\end{figure}

We partially addressed this question in Sec.~\ref{sec:method_viability}, as we established that chaotic resetting is an effective strategy to reduce the search time for angles with $CV>1$. For a given angle, we showed how $CV$ evolves with increasing uncertainty. However, we now pose the complementary question: for a given uncertainty, how many angles exhibit $CV>1$? In Fig.~\ref{fig:6}, we show $CV$ for each angle, considering uncertainties (a) $\delta=10^{-4}$ and (b) $10^{-2}$. As expected, $CV$ remains zero for angles close to the exits ($\theta_0=0$ and $\theta_0=2\pi/3$) and for angles that hit the potential walls almost perpendicularly and escape immediately ($\theta_0=\pi/3$). For $\delta=10^{-4}$, this also holds for other values of $\theta_0$, as $\delta$ is small enough to keep the initial condition within one of the smooth regions embedded in the fractal phase space. In both panels, we observe an accumulation of $CV$ values around 1. This reflects the fact that in a fully fractalized region of phase space, the system behaves locally as hyperbolic, as previously discussed.

\begin{figure}[ht]
	\centering
    \includegraphics[clip,width=0.45\textwidth,trim=0cm 0cm 0cm 0cm]{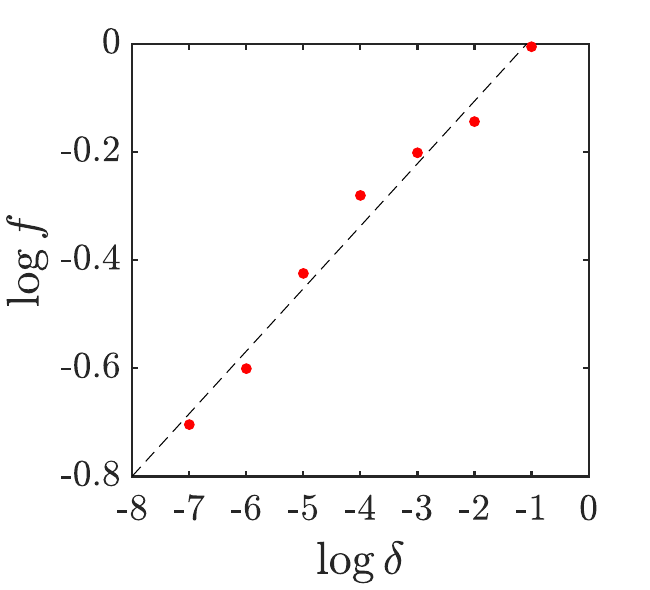}	
	\caption{Log-log plot showing the increase of the fraction of initial conditions with $CV>1$ as the uncertainty increases. The straight line suggests a relation between the variables of the form $f\propto \delta^\alpha$, where $\alpha$ is a positive constant. This indicates that the larger the uncertainty, the higher the probability that CR is a good strategy, regardless of the initial conditions. This is because it is more probable that the $\delta$-neighborhood includes a mixed phase space. From this data, we have calculated $\alpha=0.116(7)$ with a linear correlation coefficient $r=0.990$. The logarithms have base 10.}  
	\label{fig:7}
\end{figure}

Now, we compute the fraction of initial conditions with $CV>1$, $f$, for different values of the uncertainty. The results are shown in Fig.~\ref{fig:7}. As can be seen, the larger the uncertainty is, the larger the number of initial conditions for which resetting is beneficial. For increasing values of uncertainty, it is more probable to reach the boundary of a region that includes radically different escape times, either larger or smaller, thereby raising the standard deviation and increasing $CV$. In other words, it is more probable that the $\delta$-neighborhood lies within a mixed phase space. The results shown in Fig.~\ref{fig:7} suggest that the fraction of initial conditions with $CV>1$ scales with uncertainty following a power law. This behavior is expected in any chaotic system with a mixed phase space. This dependence is not surprising, as the fraction of uncertain initial conditions scales with uncertainty according to a similar scaling law, as established by Grebogi et~al. \cite{Grebogi1983}. Indeed, an initial condition will have $CV>1$ only if it is uncertain.

\FloatBarrier

\subsection{Minimizing the escape time}

Once resetting is determined to be beneficial, the remaining question is: what is the impact on the search time? For this purpose, we measure the reduction of the average escape time using the optimal resetting time $t_r^*$: $R_T(t_r^*)=(\bar{T}-\bar{T}_r)/\bar{T}$, where $\bar{T}$ the average escape time without resetting and $\bar{T}_r$ the average escape time with resetting. 

\begin{figure}[ht]
	\centering
    \includegraphics[clip,width=0.45\textwidth,trim=0cm 0cm 0cm 0cm]{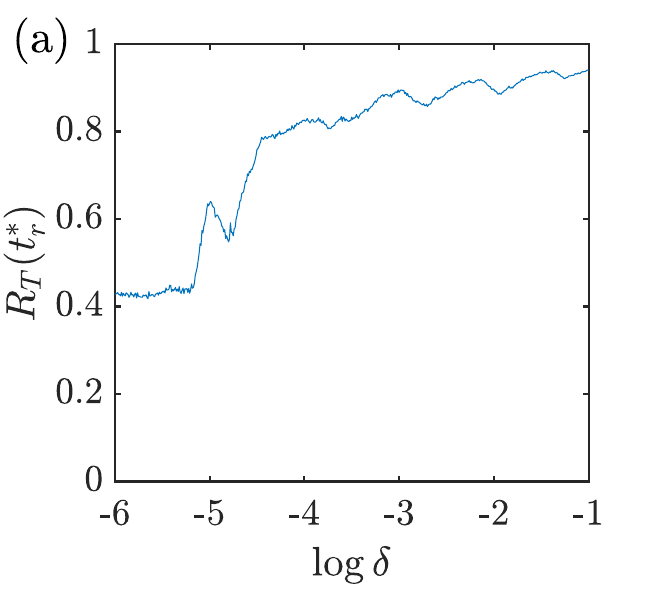}	
    \includegraphics[clip,width=0.45\textwidth,trim=0cm 0cm 0cm 0cm]{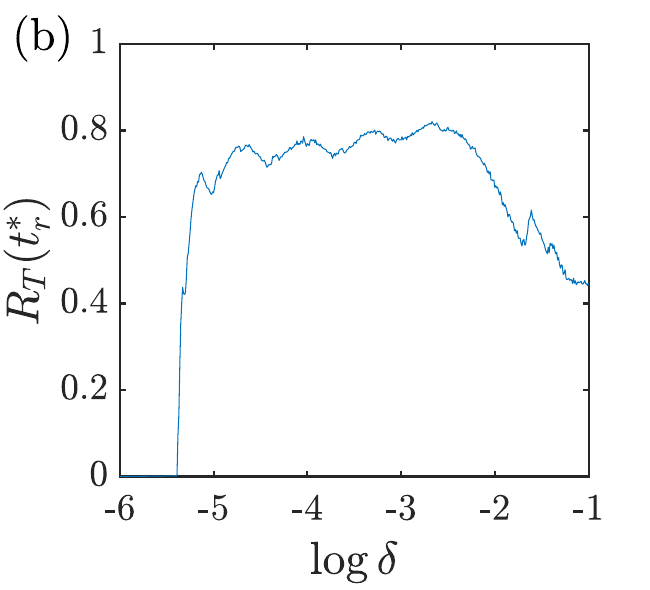}	
	\caption{Reduction in the average escape time $R_T$ obtained using the optimal resetting interval $t_r^*$, as a function of the base-10 logarithm of the uncertainty.Two representative cases are shown: (a) a chaotic trajectory and (b) a regular trajectory. The values of $\theta_0$ for each panel are the same as the ones used in Fig.~\ref{fig:3}, to facilitate direct comparison. The reduction is remarkable in every scenario except for the case in panel (b) where the $\delta$-neighborhood is small and lies entirely within a smooth region of phase space, so that all trajectories already escape fast without resetting. In practical terms, this indicates that chaotic resetting is always beneficial if a trajectory takes a period longer than the characteristic time of the ones that escape fast. }  
	\label{fig:8}
\end{figure}

We show in Fig.~\ref{fig:8} this reduction for the same angles as in Fig.~\ref{fig:3}: in panel (a) the initial condition belongs to a fractal region and in panel (b) the initial condition belongs to a smooth region. Remarkably, the reduction of the search time is significant even in Fig.~\ref{fig:8}(a) for the uncertainty values for which $CV$ is only slightly larger than $1$. This is because the access to the small smooth regions embedded in the fractal regions are sufficient to make resetting advantageous, with a reduction larger than a $40\%$. The reduction in escape time shows a peak for $\delta \approx 10^{-5}$, that is, when large smooth regions are reached. In Fig.~\ref{fig:8}(b), the initial condition belongs to a smooth region and there is no reduction on the search time until the uncertainty is larger than the width of the basin. In any case, resetting would not be necessary in this situation as the escape time is already small.

We conclude that, in practical decision-making scenarios where a detailed analysis of the underlying dynamics is not feasible or available, a preliminary exploration of the system’s behavior is sufficient to assess the suitability of chaotic resetting. In particular, for an experiment where the initial condition is affected by a certain uncertainty and we observe a variability in completion times, estimating the characteristic time scale associated with rapid trajectories provides a reliable criterion. If a given trajectory persists significantly longer than this fast time scale without completing the escape (or search) process, then implementing chaotic resetting is expected to be advantageous.

\FloatBarrier

\section{Conclusions and Discussion}
This work establishes that resetting strategies can accelerate processes in deterministic chaotic systems, not only stochastic ones. While stochastic resetting exploits the randomness inherent to diffusive processes, chaotic resetting exploits the equally fundamental but deterministic sensitivity to initial conditions. This conceptual extension bridges two previously distinct domains—stochastic optimization and deterministic chaos—and opens new avenues for process acceleration in chaotic systems.

We have illustrated this method using a chaotic scattering problem in which particles escape from a two-dimensional potential. We have identified that the essential requirement for chaotic resetting to be effective is that the system must exhibit a mixed phase space where regular and chaotic regions coexist. This ensures that resetting can substitute slow chaotic orbits by faster regular ones. This condition is not system-specific, suggesting that chaotic resetting might be applicable to reduce transient times across a broad range of problems and systems. For instance, it could reduce first passage times in multi-well potentials; shorten transients in dynamical systems such as the forced and magnetic pendulum, the Lorenz system, or population models; decrease recurrence times in Hamiltonian maps; and accelerate pattern formation in reaction-diffusion models or systems of active particles. 

Traditional approaches to chaos control focus on stabilizing unstable periodic orbits, actively controlling trajectories or suppressing chaos entirely. Chaotic resetting takes a fundamentally different approach: it accepts and exploits chaos rather than fighting it. By systematically interrupting unfavorable trajectories and leveraging the natural variability arising from sensitivity to initial conditions, we can achieve dramatic reductions in completion times (from $40\%$ to $90\%$ in the Hénon--Heiles system) without requiring detailed knowledge of the system's dynamics or precise control over initial conditions. This makes the method particularly valuable in experimental settings where such control is limited. In fact, from an experimental standpoint, the decision of whether to apply resetting can be based on a simple and practical criterion: one should assess the variability of completion times across repeated realizations of the same experiment. A large dispersion in completion times indicates that the $\delta$-neighborhood (the region of phase space affected by the uncertainty) lies within a mixed phase space. Under these circumstances, resetting should be applied to trajectories whose completion times significantly exceed the characteristic time scale associated with rapidly completing trajectories.

In summary, chaotic resetting represents not merely a technical tool for a specific problem, but a conceptual framework for accelerating deterministic chaotic processes by strategically exploiting their defining characteristic: sensitivity to initial conditions. \\

 \textbf{CRediT authorship contribution statement}

\textbf{Julia Cantisán:} Conceptualization, Methodology, Software, Visualization, Writing – original draft. \textbf{Alexandre R. Nieto:} Conceptualization, Methodology, Software, Visualization, Writing – original draft. \textbf{Jesús M. Seoane:} Conceptualization, Methodology, Writing – review \& editing.\\

 \textbf{Declaration of competing interest}

The authors declare that they have no known competing financial interests or personal relationships that could have appeared to influence the work reported in this paper.\\

 \textbf{Acknowledgments}
 
This work has been financially supported by MCIN/AEI/10.13039/501100011033 and by “ERDF A way of making Europe” (Grant No. PID2023-148160NB-I00).\\

\textbf{Data availability}

The data that support the findings are available from the corresponding author upon reasonable request.

\section*{Appendix A}

In Algorithm 1, we present the pseudocode explaining the steps to calculate the escape time of one particle using chaotic resetting. In this work, we have used the following parameter values: $h=0.001$, $E=0.17$, and $t_{max}=10^4$. Here, the escape condition is based on the position of the particle. In particular, we consider that the particle has escaped if its position exceeds a circle of radius $2$. Any circle of radius greater than $1$ will suffice, since the chaotic saddles of the potential fall in this circumference.  The same scheme can be applied to different problems, simply changing the break condition.

\begin{algorithm}[h]
\caption{Escape time of one particle using chaotic resetting. }

\KwIn{$E$, $t_r$, $\delta$, $h$, $\theta_0$, $t_{max}$}
\KwOut{$t$  \quad // {\ttfamily\small Total escape time of the process}}  


$t_{last} = 0$; \quad // {\ttfamily\small Time of the last reset}

$x = 0$; $y = 0$;
$\dot{x} = \sqrt{2E}\sin\theta_0$;
$\dot{y} = \sqrt{2E}\cos\theta_0$; \quad // {\ttfamily\small Initial conditions}
\SetKw{to}{to}\\
\SetKw{step}{step}

\For{($t=0$ \kern 0.5pc \to \kern 0.5pc $t_{max}$\kern 0.5pc \step \kern 0.5pc $h$) }{

    $(x,y,\dot{x},\dot{y}) = \text{RK4}(x,y,\dot{x},\dot{y},h)$ \quad 
    // {\ttfamily\small Runge-Kutta 4 numerical step}

        \If{particle escapes}{
        
        \SetKw{Break}{break}
        \Break\
    }

    \If{$(t - t_{last}) \ge t_r$}{
        // {\ttfamily\small Apply resetting}

        $\theta = \theta_0 + \delta\mathcal{U}(-1,1)$; \quad
        // {\ttfamily\small Perturb angle considering uncertainty}

        $x = 0$; $y = 0$;
        $\dot{x} = \sqrt{2E}\sin\theta$;
        $\dot{y} = \sqrt{2E}\cos\theta$;

        $t_{last} = t$; \quad // {\ttfamily\small Update last resetting time}
    }
}

\end{algorithm}

\FloatBarrier

%

\end{document}